\documentstyle[12pt,aasms4]{article}

\tighten

\begin{document}

\title{NUCLEAR BAR, STAR FORMATION AND GAS FUELING IN THE ACTIVE GALAXY
NGC 4303}

\vspace{0.5in}

\author{Luis Colina\altaffilmark{1,2}}
\affil{\em Space Telescope Science Institute, 3700 San Martin
Drive, Baltimore, MD21218, USA}

\and

\author{Keiichi Wada}
\affil{\em National Astronomical Observatory, Mitaka, 181-8588, Japan
(wada.keiichi@nao.ac.jp)}

\scriptsize

\altaffiltext{1}{Affiliated with the Astrophysics Division, Space
Science Department of
 ESA}
\altaffiltext{2}{Present address: Instituto de F\'{\i}sica de
Cantabria, Facultad de Ciencias,
39005 Santander, Spain (colina@ifca.unican.es)}


\normalsize

\begin{abstract}

A combination of Hubble Space Telescope (HST) WFPC2 and NICMOS images
are used to investigate the gas/dust and stellar structure inside the
central 300 pc of the nearby active galaxy NGC 4303.

The NICMOS H-band (F160W) image reveals a bright core and a nuclear
elongated
bar-like structure of 250 pc in diameter. The bar is centered on
the bright core, and its major axis is oriented in proyection along the
spin axis
of the nuclear gaseous rotating disk recently detected (Colina \&
Arribas 1999).

The $V-H$ (F606W $-$ F160W) image reveals a complex gas/dust
distribution
with a two-arm spiral structure of about 225 pc in radius. The
southwestern arm
is traced by young star-forming knots while the northeastern arm is
detected
by the presence of dust lanes. These spirals do not have a smooth
structure but
rather they are made of smaller flocculent spirals or filament-like
structures.

The magnitudes and colors of the star-forming knots are typical of
clusters of young stars with masses of 0.5 to 1 $\times $10$^5$
M$_{\odot}$, and
ages of 5 to 25 million years.

The overall structure of the nuclear spirals as well as
the size, number and masses of the star-forming knots are explained
in the context of a massive gaseous nuclear disk subject to
self-gravitational
instabilities and to the gravitational field created by the nuclear bar.
According to the model, the gaseous disk has a mass of about
5 $\times  10^7 M _{\odot}$ inside a radius of 400 pc, the bar has
a radius of 150 pc and a pattern speed of $\sim$ 0.5 Myr$^{-1}$, and
the average mass accretion rate into the core ($R < $ 8 pc) is $\sim
0.01
M_\odot$ yr$^{-1}$ for about 80 Myr.
\end{abstract}


\section{INTRODUCTION}

The fundamental question of how active galaxies are fueled and
how their active nuclei (AGNs) are nourished is still largely unsolved
despite
important theoretical efforts.
Redistribution of angular momentum of the interstellar gas
due to a stellar bar and subsequent gravitational instability of the gas
is
suggested to be a likely mechanism to fuel AGNs (Shlosman, Begelman \&
Frank
1989).
Although numerical simulations of the effects of bars in spiral galaxies
support the bar-induced fueling scenario (van Albada \& Roberts 1981;
Combes \& Gerin 1985; Athanassoula 1992; Friedli \& Benz 1993;
Shlosman \& Noguchi 1993; Heller \& Shlosman 1994;
Wada \& Habe 1995; Fukuda, Wada \& Habe
1998), fueling processes inside $\sim $ 100 pc from an AGN are still
unclear.

Observationally, there have been several authors investigating the
structure
of the nuclear regions of nearby spirals with the highest spatial
resolution
permitted from the ground (Knapen et al. 1995a,b; Elmegreen, Chromey \&
Warren 1998;
Laine et al. 1999; Ryder \& Knapen 1999). However, it is
the combination of multiwavelength HST imaging with state-of-the-art
multi-phase hydrodynamical codes that provides a unique tool to study in
detail
the gas channeling processes in the central regions of galaxies and the
fueling
of AGNs, as well as the nuclear star-formation processes, their
evolution and
connection with the AGN phenomenon.

The high spatial resolution provided by the HST cameras is crucial to
investigate the structure of the stellar, gas and dust component in the
nuclear
regions of nearby active galaxies with a linear resolution of a few
parsecs.
The ultraviolet continuum is a direct tracer for the presence of
unobscured clusters of young massive hot stars, i.e. ongoing
star-formation
processes, while the near-infrared image traces the old stellar
component, and
could reveal the presence of nuclear stellar bars. Finally, color images
(i.e.
$V-H$ for example) uncover the overall structure of the
gas/dust distribution in these regions.

NGC 4303 is a barred spiral classified as SAB(rs)bc (de Vaucouleurs et
al. 1991)
and located in the Virgo cluster (adopted distance of 16.1 Mpc
hereinafter).
This galaxy was known to have circumnuclear star formation (Pogge 1989).
HST ultraviolet images did show the presence of an unresolved UV-bright
core
connected with a UV-bright one-arm star-forming spiral of about 225 pc
in
radius (Colina et al. 1997; C97). Two-dimensional optical fiber
spectroscopy (Colina \& Arribas 1999; CA99) has shown that the UV-bright
unresolved core has the optical emission line ratios characteristic of
low-excitation Seyfert 2 galaxies or [OI]-weak LINERs while the overall
ionization is dominated by the star-forming spiral. Also, the velocity
field
of the ionized gas is consistent with the existence of a gaseous
rotating
nuclear disk of 300 pc in radius where the star-forming spiral is
embedded
(CA99).

In this paper we use HST multiwavelength imaging to investigate the
nuclear
region of NGC 4303. The new results are presented and their implications
for the bar
induced nuclear starburst $-$ AGN fueling scenario are discussed in the
context of models of 
the evolution of a gaseous nuclear disk subject to the gravitational
perturbation of a nuclear bar.

\section{OBSERVATIONS AND DATA REDUCTION}

The HST images of NGC 4303 obtained with the WFPC2 (ultraviolet and
optical) and
NICMOS (near-infrared) cameras, were retrieved from the HST/STScI
archive. The ultraviolet image was taken with the F218W filter.
The main results obtained from this image, and the details regarding the
calibration process, have already been published (C97).

The optical image was obtained with the F606W filter. This
filter has a wide bandpass of $\sim$ 1500 \AA~ where all the strong
optical
emission lines are included. A fraction of the light detected through
this filter could therefore be contaminated by the warm gas emission,
its exact
proportion depending on the region. The raw images were recalibrated
using the
recommended reference files and the updated photometric
keywords. No further manipulation was done to these images.

Finally, the near-infrared image was taken with NICMOS (camera 2; NIC2)
using
the F160W filter (i.e. HST H-band). Recalibration of
the image was done at the time of dearchival using the recommended
calibration files. Since NGC 4303 is an extended
target covering the entire NIC2 field-of-view, an assessment of the
effect of the
pedestal (extra signal induced in the detectors by the operating mode of
the cameras)
 in the final calibrated image was done. As part of the tests,
the pedestal was removed using the STScI routines (written by E.
Bergeron), and the routine developed by R. van der Marel
(see NICMOS
WEB page for a detailed description of these routines and their updates).
A direct comparison of the calibrated image obtained by the
standard
calibration pipeline with the outputs of the pedestal removal routines,
indicates that the
effect of the pedestal in the NGC 4303 image is at a very low level
($<$ 1\% inside 5$^{''}$), not affecting the light distribution in the
regions
of interest. In view of these results, the output of the standard
calibration pipeline
is used in the entire analysis presented in this paper.

The final images were rotated and oriented along the standard N$-$E
astronomical
orientation using the position angle and plate-scale keywords stored in
the
headers of the images. The $V-H$ image (i.e. the F606W $-$ F160W) was
generated
by dividing the F606W image by the F160W image after recentering and
rebinning of the NICMOS image to the pixel size of the WFPC2 image.

In the rest of the paper we refer to the F218W, F606W and F160W images
as the
UV$-$, V$-$, and H$-$band images, although the bandpass of the F606W and
F160W
filters do not coincide with those of the standard filters used in
ground-based
observatories (see Holtzmann et al. 1995 and the WFPC2 and NICMOS WEB
pages for
detailed information).

The photometric measurements for the core of the galaxy were done with an 
aperture of 0.5$''$ radius. A smaller aperture (0.3$''$ radius) was used
to measure the magnitudes of the bright star-forming knots around the nucleus.
An aperture size correction has been included to compute the final photometric values
when using the 0.3$''$ radius aperture. 
The uncertainties in the final measurements of the ultraviolet 
and optical magnitudes range from about 0.05 mags for the brightest star-forming
knots to about 0.15 for the faintest knots. For the F606W measurements, the largest
contribution to the uncertainty comes from the estimation of the galaxy contribution 
to the total flux within the aperture. The galaxy contribution per pixel is computed
as the azimuthal average surface brightness at the distance of the knot from the core
of the galaxy. For the F218W measurements the largest contribution to the uncertainty
of the ultraviolet magnitudes comes from the uncertainty in the absolute calibration of 
the F218W filter.

\section{OBSERVATIONAL RESULTS}

\subsection{Nuclear Bar and Unresolved Core in NGC 4303}

The H$-$band image (Figure 1) reveals two salient features: an
elongated, bar-like structure (nuclear bar), and an unresolved bright core.
The nuclear bar is centered on the unresolved bright core of the galaxy,
has a diameter of about 250 pc and an axial ratio of 2:1. This nuclear bar
is about 3 to 6 times smaller than the inner  kiloparsec-size bars detected
in other active galaxies like NGC 4321 (Knapen et al. 1995) or NGC 1068
(Scoville et al. 1988).
The major axis of the bar in NGC 4303 is oriented along PA 40,
coincident in
projection with the spin axis of the nuclear gaseous rotating disk
recently
detected (CA99). Also, the northern tip of the bar coincides
with the beginning of the UV-bright one-arm star-forming spiral (Figure
2; see
also C97).

The H-bright core, which is also the brightest compact source in the
UV$-$ and
V$-$band,
is characterized by a maximum size of 8 pc, a visual (V) magnitude of
16.7 and
colors of $-$0.8 ($UV-V$) and $+$3.2 ($V-H$). The UV magnitude could
correspond
to that of a massive stellar cluster of about 10$^5$ M$_{\odot}$.
However the
observed colors are redder than those expected for young stellar
clusters, in the
absence of internal extinction. For
an instantaneous burst characterized by a Salpeter IMF and an upper mass
limit
of 65 M$_\odot$, simulations based on Bruzual \& Charlot spectral energy
distributions
(Bruzual \& Charlot 1993), predict $UV-V$ colors in the
range $-$2.8 to $-$0.8 for clusters 2.5 to 50 Myr old, and $V-H$ colors
from $-$0.7 to $+$1.5 for the same age range.

If absorption by dust were present,
an internal extinction equivalent to E(B-V)=0.1 would deredden the
observed $UV-V$
and $V-H$ colors by 0.7 and 0.3 mag, respectively. Therefore, small
amounts of
extinction make the observed $UV-V$ color ($-$1.5 after correction for
extinction)
compatible with the predictions for young stellar clusters. This is however
not be the case
for the observed $V-H$ color, where the large value (3.2) can not be
accounted for
by small amounts of dust extinction, and suggest the existence of an
extremely red source.
Such a source could be an AGN that, if characterized by a pure
power-law with an spectral index of 1 (i.e. F$_{\nu} \propto \nu^{-1}$)
would have
colors $UV-V$= $-$0.5 and $V-H$ $=$ 2.3, closer to the observed values
but still not fully compatible.

In summary, neither a young stellar cluster nor a standard AGN fully
explain the observed
colors and therefore the nature of the central bright unresolved
ionizing source in NGC 4303 is still
not solved. Ultraviolet and optical spectroscopy are required to detect
the presence of
an AGN and/or young stellar clusters and characterize their properties.

\subsection{The Overall Structure of the Nuclear Region in NGC 4303}

The $V-H$  image of NGC 4303 (Fig. 3) shows a complex dust/gas and
stellar
structure inside 4$^{''}$ ($\sim$ 310 pc) around the unresolved
core. The
one-arm star-forming spiral delineated by the UV-bright knots is still
visible
in the $V-H$ image.
The star-forming spiral is composed of several independent filaments
that seem
to run tangential to an imaginary circle of about 200 pc in radius. This
image
also shows a complex system of dust lanes and filaments mostly grouped
along two main spiral-like structures that become visible due to the
differential extinction they produce.

The most prominent dust lane, located northeast of the core at a
distance of
about 225 pc, has a $V-H$ color of 4.1, i.e. redder than the average
$V-H$
color (2.7 $\pm$ 0.3; de Jong \& van der Kruit 1994) of face-on spirals.

An extinction of 1.7 mags in the visual (A$_V$) is implied for this
region
if the difference between the observed $V-H$ color of the dust lane and
that of
face-on spirals is purely due to absorption. The rest of the dust lanes
are
less prominent in the $V-H$ map and therefore these regions are less
affected
by extinction.

\subsection{Masses and Ages of the Nuclear Star-Forming Knots in NGC
4303}

The ultraviolet and optical magnitudes of about a dozen UV-bright
knots located along the star-forming spiral have been measured. No
internal
extinction corrections have been applied to these magnitudes. For
comparison,
the $UV$ and $V$ magnitudes for a stellar cluster characterized by a
Salpeter IMF and
an upper mass limit of 65 M$_{\odot}$ have been computed as a function
of mass
and age for instantaneous bursts, using Bruzual \& Charlot spectral
energy
distributions (Bruzual \& Charlot 1993). The results of the observed and
modeled
magnitudes and colors are presented in Figure 4.

The typical mass of the star-forming knots lies in the 0.5 to 1.0 $\times 10^5
M_{\odot}$ mass range
although some have smaller (0.3 $\times 10^5 M_{\odot}$) or larger (2.3
$\times
10^5 M_{\odot}$) masses. These star-forming knots can be
separated into two well defined age groups. The first group corresponds
to
knots of 5.0 to 7.5 Myr and the second group to knots that
are 10 to 25 Myr old. Moreover, this age segregation
seems to be associated with position with respect to the nuclear bar.
All but one
of the knots located to the east of the bar have ages older than 10 Myr
while
all knots located western of the bar are 7.5 Myr, or younger. This
result is
also supported by images of the ionized gas distribution (see Fig. 3 in
CA99)
where the dominant H$\alpha$ emitting regions are associated with the
star-forming
knots located on the west side of the bar.

Internal extinction could affect the masses and ages of the star-forming
knots derived
above. As shown in Figure 4, correcting the observed magnitudes and
colors by an extinction
of E(B$-$V)= 0.1, increases the mass of the knots by a factor of
about 2,
shifting the ages from 5$-$25 Myrs to 2.5$-$7.5 Myrs, i.e. towards
younger stellar clusters.
Thus, if extinction in these star-forming regions, the masses and ages 
derived above represent a lower limit to 
the masses and ages of the stellar clusters.

\section{NUMERICAL HYDRODYNAMICAL MODELING}

\newcommand{\beq}{\begin{equation} }
\newcommand{\eeq}{\end{equation}}
\newcommand{\bea}{\begin{eqnarray} }
\newcommand{\eea}{\end{eqnarray}}
\def\mbf#1{\mbox{\boldmath ${#1}$}}

\subsection{The Numerical Method}
Previous models of the gaseous response in a non-axisymmetric, rotating
potential
assumed that the interstellar medium (ISM) is a {\it single phase}
fluid, which is unrealistic for the conditions prevailing in the nuclear
regions
(i.e. inner 1-2 kpc) of spiral galaxies.  A newly developed
hydrodynamical code
(Wada \& Norman 1999)
represents the multi-phase structure of the interstellar medium on a kpc
scale with a few pc resolution, taking into account the self-gravity of
the gas, radiative cooling and heating processes. This is a great
advantage comparing to the single-phase isothermal gas models, because
the formation and evolution of low temperature ($<$ 100 K), high
density ($\gg$ 100 M$_\odot$ pc$^{-2}$) star-forming sites can be
investigated. The model presented here does not
include the effects of star formation explicitly, such as energy
feedback from
supernova explosions and stellar winds due to massive stars.
 Those effects have been implemented in our code, and will be discussed
in a subsequent paper.

We use the second-order Euler mesh code to solve the hydrodynamical
equations, which
is based on the Advection Upstream Splitting Method) (AUSM; Liou \& Steffen 1993).
AUSM is remarkably simple, but
it is accurate enough based on comparisons with the
 Flux-difference splitting scheme (e.g.
Roe splitting) and the Piecewise Parabolic Method (PPM; Woodward \& Colella 1984).
Radiative cooling is implemented using an
implicit method with a time-independent cooling function (Spaans \& Norman 1997).
With this code, we can deal with more than
seven orders of magnitude in density and temperature.
The numerical code is tested for various 1-D and 2-D test problems.
Details of the numerical scheme and the test results
are described elsewhere (see also \cite{WN99}).

Here we show one test calculation: resonant excited spirals in a
rotating bar
potential.
The gaseous response to the non-axisymmetric potential
has been well studied numerically (e.g. Athanassoula 1992; Wada \& Habe
1992, 1995; Englmaier \& Gerhard 1997), and analytically
(e.g. \cite{YC89}; \cite{WA94}).
A rotating bar-like potential causes resonances in orbits, and
the gas forms spiral-like shocks.
The model used here is the same as model B in Fukuda, Wada, \& Habe
(1998),
in which an isothermal equation of state is assumed with the Smooth
Particle
Hydrodynamics (SPH) method.
In this model, a nuclear mass concentration (e.g. a supermassive
black hole) causes a new Lindblad resonance inside an ordinary inner
ILR.
This resonance has similar characteristic as the outer ILR, and
therefore two {\it trailing spirals} are generated near the resonance,
as also shown by analytical studies (\cite{WA94}).
Figure 5 represents the surface density distribution of the model at
$t=0.2$ ($256^2$ meshes are used).
As expected, two trailing spiral shocks are formed, and the structure
is consistent with the SPH model (see Fig.5 in Fukuda et al. 1998).

\subsection{The Hydrodynamical Model}

We use $512^2$ Cartesian
grid points for a $(800 {\rm pc})^2 $ region,
therefore the spatial resolution is 1.56 pc.
The gravitational potential of the gas is solved by
 the convolution method with FFT using $1024^2$ grid cells.

In order to investigate the effects of the stellar bar on the global and
local gaseous dynamics,
evolution of the rotating disk is simulated in an external
fixed potential $\Phi_{\rm ext}$.
We assume a time-independent external potential $\Phi_{\rm ext} \equiv
\Phi_{\rm BH} + \Phi_{\rm bar} + \Phi_{\rm disk}$, where $\Phi_{\rm
BH}$,
$\Phi_{\rm bar}$, and $\Phi_{\rm disk}$ are the potential of
a central massive black hole, stellar bar, and extended disk
potential, and defined as
$\Phi_{\rm BH} \equiv -GM_{\rm BH}/(R^2+ a_b^2)^{1/2} $,
with $M_{\rm BH} =10^7 M_\odot$ and $a_b=10$ pc,
$\Phi_{\rm disk} \propto v_d^2/(R^2+ a_d^2)^{1/2} $,
with $a_d=3$ kpc and $v_d= 200$ km s$^{-1}$,
$\Phi_{\rm bar} \propto v_c^2 /(R^2+ a^2)^{1/2}[1+ \varepsilon_0
(1+aR^2)/(R^2+a^2)^{3/2}\cos2\phi] $
with $a = 150$ pc and $v_c=80$ km s$^{-1}$.
The potential model is based on the model used in
Wada, Sakamoto, \& Minezaki (1998).
The ionized gas velocity field (CA99) has been used to
determine $a,a_d,v_c$ and $v_d$ while
$\varepsilon_0$ is a free parameter that represents the strength of the
bar.
We assume $\Phi_{\rm bar}$ rotates rigidly with a pattern speed
$\Omega_p$,
which is another free parameter.

The initial gas disk is axisymmetric and rotationally supported with a
surface density distribution given by $\Sigma_g(R) \propto
\exp(-R/R_0)$,
where $R_0 = 50$ pc.
The initial temperature of the gas is set to $10^4$ K in the whole
region.
Random density
and temperature fluctuations are added to the initial disk, which
are less than 1\% of the unperturbed values.

The free parameters of the hydrodynamical model, i.e. the
strength of the bar ($\varepsilon_0$), the pattern speed of the bar
($\Omega_p$), and the total mass of gas ($M_g$), are constrained by the
main characteristics of the nuclear region: (1) the spiral structure
located at
a radius of $\sim$ 225 pc (hereafter we refer as {\it global spiral})
cannot be traced by a smooth spiral, but
made up of smaller `flocculent' spirals or filament-like structures, and
(2) about a dozen of young (age $\sim$ 5 $-$ 25 Myr), compact (size
$\sim$ 10 pc),
 and massive ($\sim$ 0.5 $-$ 1 $\times  10^5 M_{\odot}$) star-forming
knots have been identified in the spiral structure. We postulate that
the observed massive star clusters are formed from dense compact
molecular clouds which are resolved as dense and cold gas clumps
in our simulations. In this picture the mass of the gaseous clumps must
be greater than the mass of the observed stellar clusters.

\subsection{Numerical Results}

\subsubsection{Properties of the Nuclear Stellar Bar and Gaseous Disk}

The quasi-stable configuration of the best model
fitting the observational constraints is presented in Figure 6. In this
model the bar is characterized by a strength $\varepsilon_0 =0.15$,
and a pattern speed $\Omega_p = 0.5$ Myr$^{-1}$, and the total mass of
gas within a radius of 400 pc is $M_g = 5 \times 10^7 M_\odot$.

As observed in the $V-H$ image (Figure 3), the modeled global spirals
are made of many small flocculent spirals or filament-like structures
with
small scale high density regions ($\Sigma_g \sim 100 M_\odot$ pc$^{-2}$)
and very high density clumps ($\Sigma_g > 1000 M_\odot$ pc$^{-2}$).
These high density gaseous clumps or filaments
 have temperatures in the 10 $-$ 1000 K
range, the Jeans length is smaller than the spatial resolution,
and we can expect star-formation taking place in these clumps.
These gaseous clumps could therefore be the
progenitors of the observed massive star-forming knots.

On the other hand, the resultant gaseous morphology would be very
different from the observed one if the bar were rotating with a slow
pattern velocity,
e.g. $\Omega_p = 0.2$ Myr$^{-1}$. The two global spirals would be formed
at 
larger radii (between 300 and 600 pc) while the detected spiral structure
in
NGC 4303 is located at a radius of 225 pc.
Therefore, the pattern speed of the nuclear bar in this galaxy
rotates much faster than $\Omega_p = 0.2$ Myr$^{-1}$, and it rotates
probably
at a nearly maximum rate, i.e. $\Omega_p = 0.5$ Myr$^{-1}$.
Consequently the two global spirals observed in the $V-H$ image are
caused by the outer Lindblad resonance produced by the nuclear bar.
To determine the pattern speed more precisely we would need a comparison
between models and  higher spatial resolution kinematical information of
the
molecular and ionized gas. Our conclusion at this moment is that the
maximum
pattern speed of the bar (i.e. $R_{\rm bar} = R_{\rm CR}$) is
consistent with the observed distribution of clusters of young massive
stars.

For more quantitative argument on the morphology of the dense gas,
 in Fig.7, we plot the global Fourier Amplitude
of dense regions against the threshold density above which
`clouds' are identified.
The global Fourier Amplitude is defined by
\bea
 |C_m(\Sigma_t)| \equiv \frac{
\left| \int_{\Sigma_g > \Sigma_t} \Sigma_g(\mbf{x}) e^{-im \phi}
d\mbf{x} \right|
 }{\int_{\Sigma_g > \Sigma_t} \Sigma_g(\mbf{x}) d\mbf{x} },
\eea
where $\Sigma_g$ is the gaseous surface density and $\Sigma_t$ is
the density threshold . We plot $|C_1(\Sigma_t)|$  and $|C_2(\Sigma_t)|$
for models $A$ (the best-fiting model), $B$ and $C$ with
$[\varepsilon_0,
 M_g (10^8 M_\odot)]$ $ = (0.15,0.5)$,
 (0.15, 0.1) and (0.05, 0.5), respectively.
The pattern speed of the bar is $\Omega_p = 0.5$ Myr$^{-1}$ in
all models.
Larger $|C_2|$ means that the dense clouds are distributed more
bi-symmetrically, in other words,
 the two-armed spiral structure is more clearly seen.
For the higher threshold density, the number of clouds becomes small,
and more compact clouds are identified.
 The positive correlation between $\Sigma_t$ and
$|C_1|$ or $|C_2|$ in all models means that
the denser clumps
are distributed more non-axisymmetrically.
In all three models,
$m=1$ and $m=2$ modes are comparable for
the entire range of $\Sigma_t$.
This is consistent with the structure
of the high density regions ($> 10^3 M_\odot$ pc$^{-2}$)
not being the same in the two global spirals.
This could explain why the young star clusters of NGC 4303
are clearly seen only in one of the spiral arms.
This is an interesting subject to
verify by future simulations in which star formation processes are
explicitly implemented.

Figure 7 also shows that the one and two-arm spiral structures are
less prominent in model $C$ than in model $A$ and $B$ by an order of
magnitude. This is simply because
the non-axisymmetric component of the potential is weaker in model $C$.
We do not observe prominent global spiral patterns in model $C$.
The clumps are distributed almost randomly, and less dense filaments
form multi-arm structure.

We therefore conclude that a bar characterized by a very weak potential
($\varepsilon_0 \sim 0.05$), or a weak bar (the axial ratio $\sim
0.8-0.9$),
cannot reproduce the observed {\it global spiral pattern}, although
many spirals/filaments on a small scale ($\sim 10-100$ pc) are formed.
In this case possible star forming sites (i.e. dense and cold clumps)
would be
randomly distributed over the entire disk, contrary to what is observed.
This is due to the perturbation from the nuclear bar being
too weak to overwhelm random perturbations due to the local
self-gravitational instability.
With $\varepsilon =$ 0.15 and 0.2, on the other hand,
a two-armed spiral enhancement
in a quasi-stable multi-phase gas disk is formed, although
the difference between results for  $\varepsilon_0=0.15$ and 0.2 is not
significant.

We next evaluate the total mass in the central region of NGC 4303.
We compare two models with $M_g =$  1 and 5 $\times 10^7 M_\odot$.
 A more massive disk $(M_g> 10^8 M_\odot)$
can be excluded, because such disk is highly unstable against
gravitational instabilities, and would have fragmented into too
many ($N \lesssim 100$) discrete clouds contrary to what is observed.
The morphology of the gas disk in a quasi-steady state in
model $B$ $(M_g=1 \times 10^7 M_\odot)$ is similar
to the model $A$ $(M_g=5 \times 10^7 M_\odot)$.
However the mass of the cloud is different.
As we discussed in \S 3,
about a dozen young star forming knots are observed in the
central region, 600 pc in size, of NGC 4303.
Since these clusters are young ($\sim 5$ Myr) and
compact ($< 10$ pc), we suspect that the progenitors of these star
clusters are  compact ($\sim 10$ pc) gas clouds.
We identify clouds for a given density threshold and $T<100$ K
from the numerical data, and summarize their properties in Table 1.
Since the mass of the observed star clusters is $\sim 0.5-1.0 \times 10^5 M_\odot$,
the gas clouds in model $B$ are apparently too light, if
the star formation efficiency is about 0.1.
Therefore we conclude that the total gas mass inside $R=400$ pc of
NGC 4303 should be about $5\times 10^7 M_\odot$, not $1\times 10^7
M_\odot$. From the observed rotational velocity ($\sim 97$ km s$^{-1}$ at $R=350$
pc),
the dynamical mass in this region is estimated as about $8\times10^8
M_\odot$.
That is, the gas mass fraction to the total mass is about 6\%.

\subsubsection{Mass Accretion Rate into the Core}

Finally, the best fitting model predicts that gas would be accreted into
the unresolved bright core ($R<$ 8 pc).
The time evolution of the total gas mass inside radii
$R = 4,8,12,16,20$ and 24 pc is represented in Figure 8.
The average accretion rate into the observed core ($R<$ 8 pc) is
$8.8 \times 10^{-3}$ $M_\odot$ yr$^{-1}$ from $t=20$ to $t=100$ Myr.
If the accreted gas is involved in a star formation process,
the total mass of stars formed during this period would be
$\sim 7\times 10^4 M_\odot$ provided that the star formation
efficiency is $\sim 0.1$. This is consistent with the
estimated mass of the nuclear stellar cluster if the ultraviolet 
luminosity of the unresolved UV-bright core were entirely 
emitted by young massive stars (\S 3.1).
On the other hand, if a fraction of the accreted mass feeds
a central supermassive black hole, the emitted energy would dominate
the radiative energy of the core. However, with
our spatial resolution ($\sim 1.6$ pc),  it is still unclear
the gas accretion process and rate in the core.

\section{CONCLUSIONS}

Hubble Space Telescope ultraviolet, optical and near-infrared images
have been used to investigate the gas/dust and stellar structure in the inner
300 pc of the nearby active galaxy NGC 4303.
The H-band (F160W) image reveals a bright unresolved core ($<$ 8 pc) coincident 
with the ultraviolet and optical nucleus, and a
nuclear bar-like structure (nulcear bar) of 250 pc in diameter. 
The nuclear bar is centered on
the bright core, and its northern tip coincides with the beginning of
the
UV-bright one-arm star-forming spiral.

The $V-H$ (F606W $-$ F160W) image reveals a complex gas/dust
distribution
with a two-arm spiral structure of about 225 pc in radius.  These spirals 
do not have a smooth structure but
rather they are made of smaller flocculent spirals or filament-like
structures.

The ultraviolet and optical magnitudes and colors of the star-forming
knots are
typical of clusters of young stars with masses 0.5 $-$ 1 $\times $10$^5$
M$_{\odot}$,
and ages of 5 to 25 million years.

Based on our two-dimensional multi-phase hydrodynamical simulations,
in which self-gravity of the gas, radiative cooling, and
the galactic rotation are considered, we conclude that (1) the star
clusters are formed due to gravitational instability along the cold
dense gas spiral driven by the outer Lindblad resonance of the nuclear
bar
whose pattern speed is $\sim 0.5$ Myr$^{-1}$ (corotation at a radius of
150 pc), and
(2) the total gas mass inside $R<400$ pc is $\sim 0.5$-$1.0 \times 10^8
M_\odot$, which is roughly 10\% the dynamical mass ($M_{\rm dyn}$)
inferred in this region (CA99).
The average gas accretion rate into  $R < 8$ pc predicted by the best 
fitting model is about 0.01 $M_\odot$ yr$^{-1}$ for 80 Myr.

\acknowledgments We acknowledge stimulating discussions with C. Norman.
KW thanks Yamada Science Foundation for their support at STScI.
Numerical computations were carried out on
VPP300/16R at the Astronomical Data Analysis Center of the
National Astronomical Observatory, Japan.

\newpage
\vspace*{0.5cm}

\figcaption{HST F160W (H-band) image of the nuclear region of NGC 4303
 showing the bright unresolved core and the elongated bar-like structure
 (nuclear bar) of about 250 pc in size and oriented along PA 40. The first 
three contour levels represent a surface
 brightness of 16.4, 16.0 and 15.54 mags. arcsec$^{-2}$, respectively.
The rest of the contours increase in steps of 0.25 mags up to a level of
11.54 mags. arcsec$^{-2}$ for the inner most level. At an assumed
distance of 16.1 Mpc, one second of arc corresponds to a linear size of
78 parsecs. }

\figcaption{Composite image showing the nuclear region of NGC 4303 in
the
UV$-$ (grey scale) and H-band (contours). The UV-band image clearly
shows
the one-arm spiral structure of ongoing massive star formation
connected with the unresolved core of NGC 4303. The UV flux levels cover
the 0.14 $-$ 3.0 $\times$ 10$^{-15}$ erg s$^{-1}$ cm$^{-2}$ \AA$^{-1}$
arcsec$^{-2}$
range.
The H-band contours (with the
same levels as in Figure 1 starting in 16.0 mags. arcsec$^{-2}$) show
that
the northern tip of the nuclear bar coincides with the beginning
of
the UV-bright one-arm spiral.}

\figcaption{Composite image showing the nuclear region of NGC 4303 in
the
$V-H$ color (grey scale) and H-band (contours with same levels as in
Figure 2).
The V$-H$ image traces the
overall gas/dust and stellar structure showing the two-arm (star-forming
and dust lane, respectively) global spiral with its smaller flocculent
spirals
and filaments. The measured $V-H$ colors cover a range from +4.1
(dustier regions
represent in white) to +2.6 (bluer regions represented in black).}

\figcaption{Diagrams showing the measured photometric properties of the
star-forming regions (filled triangles in the diagrams) located at a
radius of about
200 pc from the unresolved core of NGC 4303. 
The uncertainties in the ultraviolet (UV) and optical (V) magnitudes range from
about 0.05 mags for the brightest knots to about 0.15 mags for the faintest,
i.e. from $\sim$ 0.07 to 0.2 for the UV $-$ V color.
The apparent ultraviolet
(F218W) and visual (F606W) magnitudes of
5$\times$ 10$^4$ M$_{\odot}$ (open squares) and 1$\times$ 10$^5$
M$_{\odot}$ (open circles)
massive star clusters at a distance of 16.1 Mpc are synthesized at
different ages
(from 1 to 100 Myr indicated by the numbers in the figures) using
Bruzual \& Charlot
models and are also representedd for comparison.}

\figcaption{A test calculation: spiral shocks produced by a resonance
due to a bar potential with a supermassive black hole.}

\figcaption{Surface density map in 400 pc$\times$ 400 pc at an age of
$t=46$ Myr
 according the hydrodynamical model that best fits the observational
constraints.
 The color bar shows the log-scaled surface density between 10$^{-1}$
and
 $10^4 M_\odot$ pc$^{-2}$. The global morphology does not change
significantly
 after $t\sim 20$ Myr until 100 Myr.}

\figcaption{Global Fourier Amplitude ($|C_m|$) of clumps
as a function of the threshold density normalized by
the mean surface density $\langle \Sigma_g\rangle$
for
model $A$ [thick solid line ($m=2$) and dotted line ($m=1$)],
model $B$ [thin solid line ($m=2$) and dot-dash line ($m=1$)], and
model $C$ [solid line with diamonds ($m=2$) and dash line ($m=1$)]  }.

\figcaption{Evolution of the total gas mass in five regions of
decreasing radius, from 24 to 4 pc, for the model shown in Figure 6. The
unit of the gas
mass is $M_\odot$. The line at $R_0 = 8$ pc shows an accretion rate,
$8.8\times
10^{-3} M_\odot$ yr$^{-1}$.}

\newpage
\begin{deluxetable}{ccccccc}
\tablewidth{33pc}
\tablecaption{Properties of Dense Clouds in Numerical Models}
\tablehead{
 & ${\Sigma_t}$\tablenotemark{a} & ${N_c}$\tablenotemark{b}  &
${M_c}$\tablenotemark{c} & ${M_{\rm c,max}}$\tablenotemark{d} &
${M_{\rm c,min}}$\tablenotemark{e} & ${L_c}$\tablenotemark{f} \\
 model  &$(M_\odot$ pc$^{-2})$ &  &$(10^5 M_\odot)$  & $(10^5 M_\odot)$
& $(10^3 M_\odot)$  & (pc)
}
\startdata
$A$\ldots\ldots  & 3000 & 20 &  1.4 & 5.8  &   7.5  & 2.9 \nl
$B$\ldots\ldots  & 400  & 19 & 0.12 & 0.53 &   0.98 & 2.6 \nl
$C$\ldots\ldots  & 4000 & 21 &  1.7 & 26.0 &   9.9  & 2.3 \nl
\tablenotetext{a}{threshold surface density}
\tablenotetext{b}{number of clouds}
\tablenotetext{c, d, e}{average, maximum, and minimum mass of the $N_c$
clouds}
\tablenotetext{f}{average size of the $N_c$ clouds}
\enddata
\end{deluxetable}

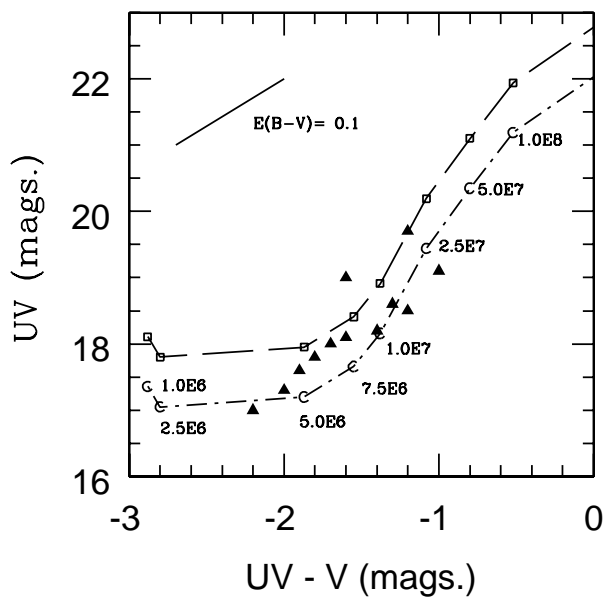
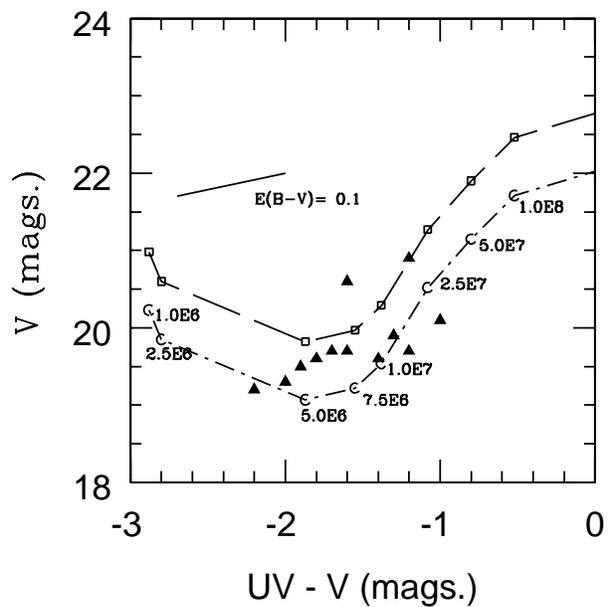
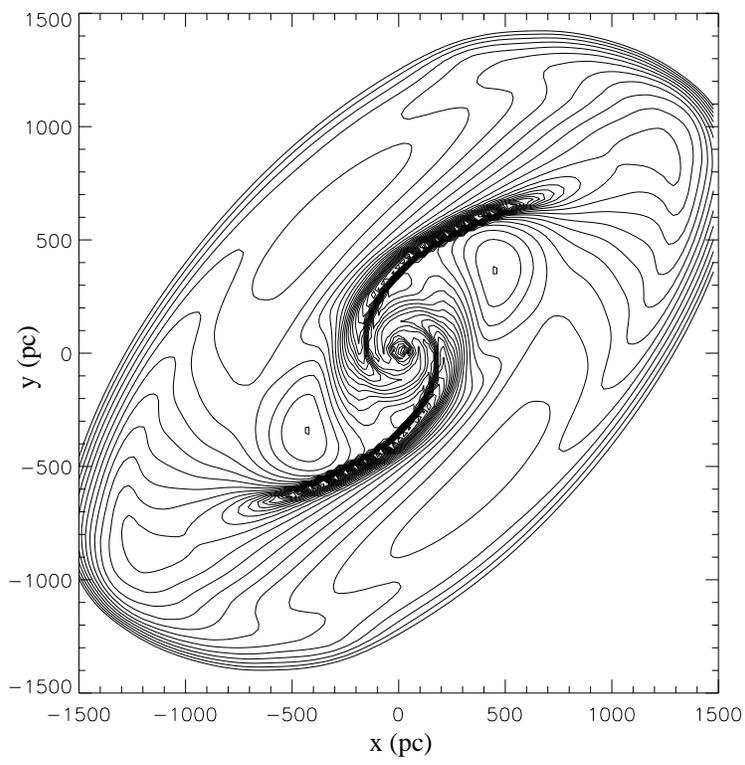
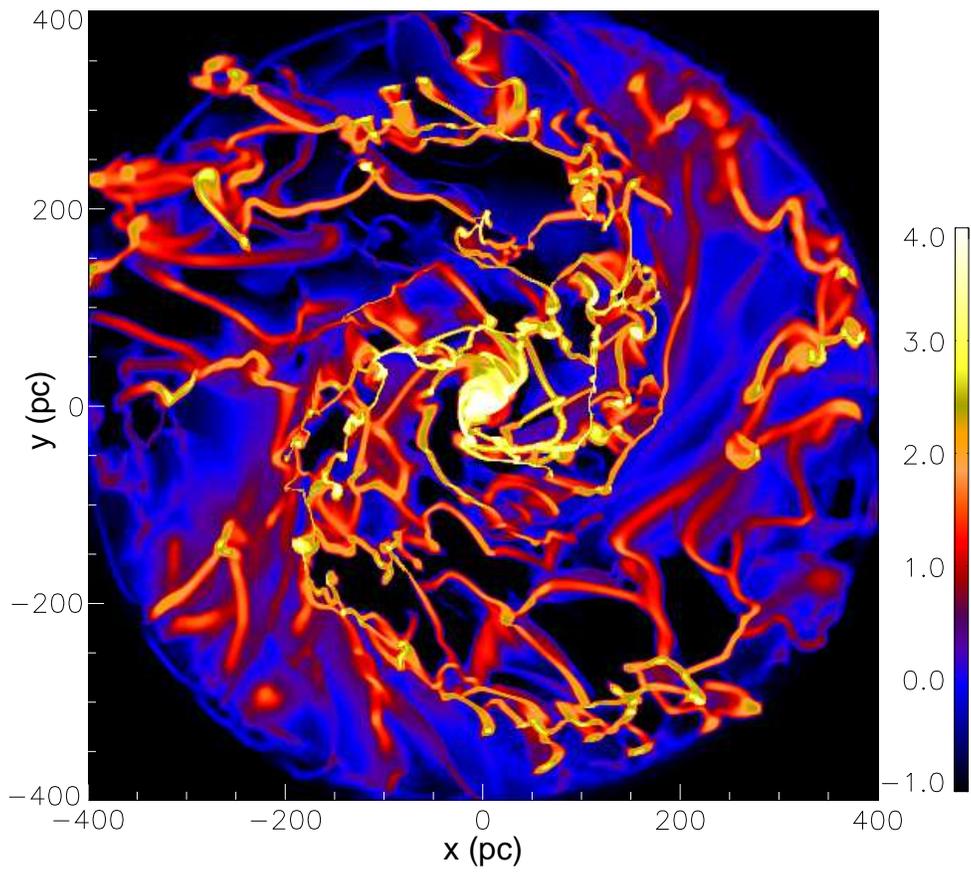
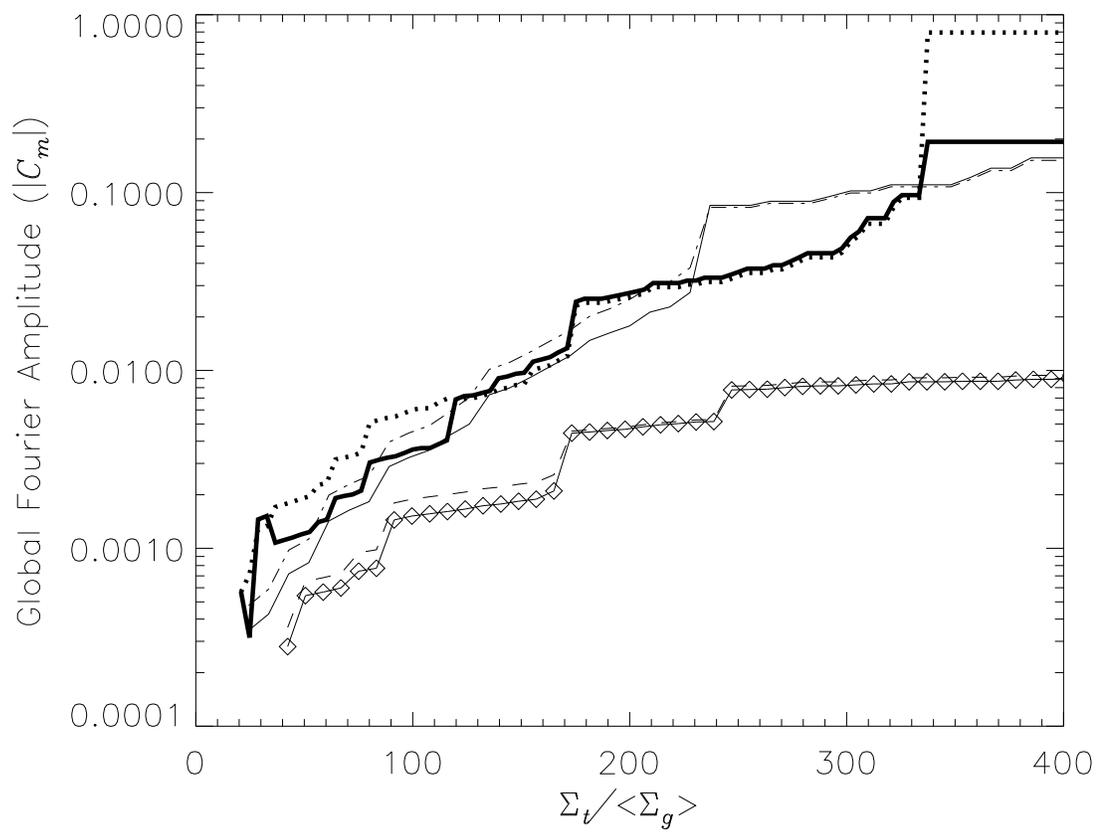
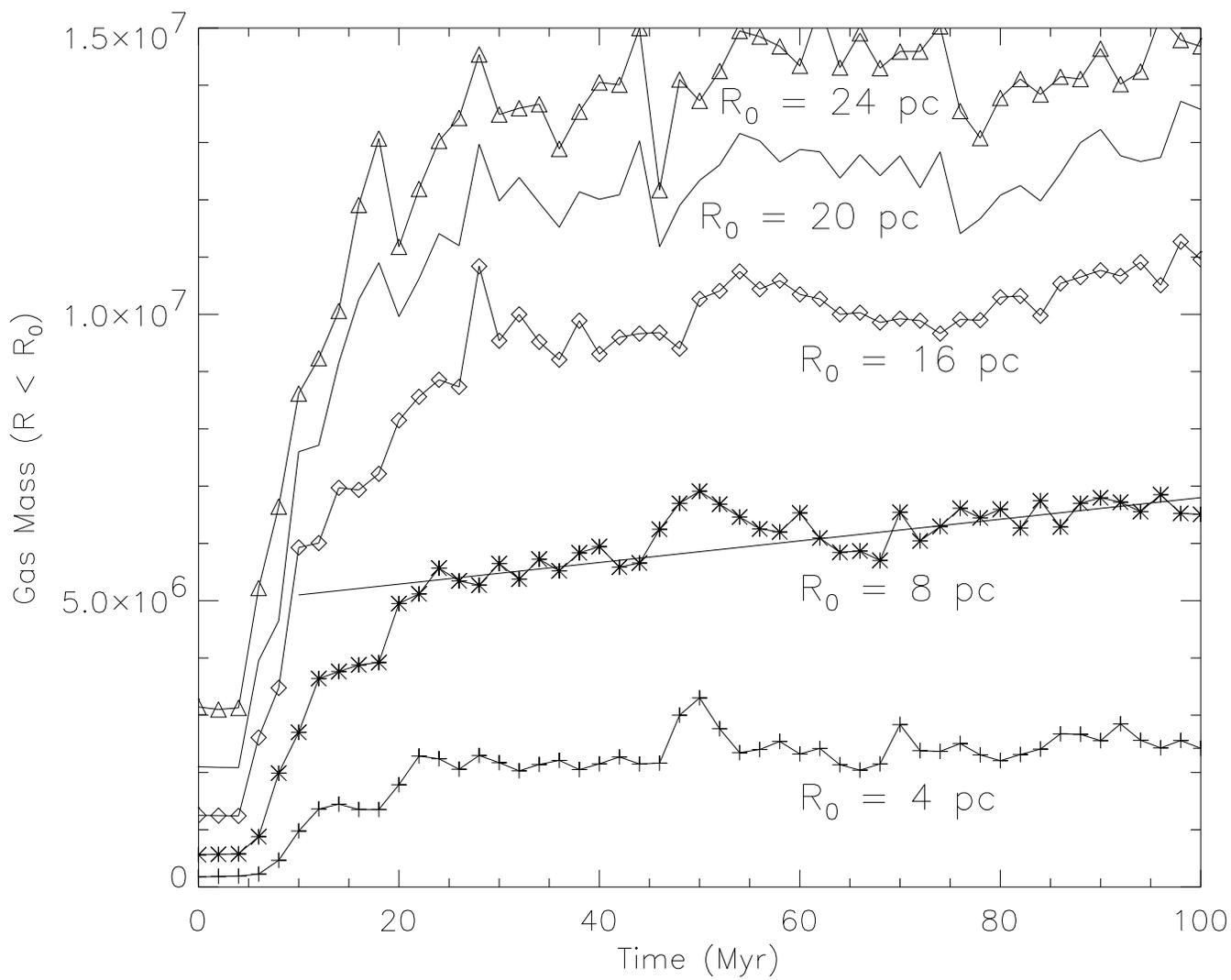
\end{document}